
\input harvmac
\input epsf

\def\Scal{{\cal S}}
\def \inparg{\leftskip = 40pt\rightskip = 40pt}
\def \outparg{\leftskip = 0 pt\rightskip = 0pt}

\def\npb{{Nucl.\ Phys.\ }{\bf B}}

\def\plb{{Phys.\ Lett.\ }{\bf B}}
\def\prd{{Phys.\ Rev.\ }{\bf D}}
\def\prl{Phys.\ Rev.\ Lett.\ }

\def\half{{\textstyle{1\over2}}} 
\def\frak#1#2{{\textstyle{{#1}\over{#2}}}}
\def\frakk#1#2{{{#1}\over{#2}}}

\def\ga{\gamma}
\def\sy{supersymmetry}
\def\sic{supersymmetric}

\def\lf{16\pi^2}

\def\GeV{{\rm GeV}}
\def\TeV{{\rm TeV}}

{\nopagenumbers
\line{\hfil CERN-PH-TH/2004-009}
\line{\hfil LTH 616}
\line{\hfil hep-ph/0402045}
\vskip .5in    
\centerline{\titlefont R-parity Violation and}  
\vskip 0.2cm
\centerline{\titlefont General Soft 
Supersymmetry Breaking}
\vskip 1in
\centerline{\bf I.~Jack, 
D.R.T.~Jones\foot{address
from Sept 1st 2003-31 Aug 2004:
TH Division, Department of Physics, 
CERN, 1211 Geneva 23, Switzerland} and A.F.~Kord}
\medskip
\centerline{\it Dept. of Mathematical Sciences,
University of Liverpool, Liverpool L69 3BX, UK}
\vskip .3in

We consider the most general class possible  of soft supersymmetry
breaking terms that can be added to the MSSM, with  and without R-parity
violation, consistent with the sole requirement that no quadratic
divergences  are induced. We renormalise the resulting theory through
one loop and  give an example of how a previously ignored term  might
affect the sparticle spectrum.

\Date{February 2004}

The minimal \sic\ standard model (MSSM) consists of a supersymmetric 
extension of the standard model, with the addition of a number  of
dimension 2 and dimension 3 \sy-breaking mass and interaction terms.  
It is well known that  the MSSM is not, in fact, 
the most general renormalisable field theory consistent 
with the requirements of gauge invariance and naturalness; 
the unbroken theory is augmented by a discrete symmetry 
($R$-parity) to forbid a set of baryon-number and lepton-number 
violating interactions, and the \sy-breaking sector omits 
both $R$-parity violating soft terms and a set 
of ``non-standard'' (NS) soft breaking terms. There is a large literature 
on the effect of R-parity violation; a recent analysis 
(with ``standard'' soft-breaking terms) and references 
appears in Ref.~\ref\ADD{B.C.~Allanach, A.~Dedes and H.K.~Dreiner,
hep-ph/0309196}; for earlier relevant work see in particular 
\ref\deCarlosDU{
B.~de Carlos and P.L.~White,
\prd 54 (1996)  3427;  {\it ibid\/} 55 (1997)  4222 
}. The need to consider NS terms  in a model--independent
analysis was stressed in
Ref.~\ref\lrlh{L.J. Hall and  L. Randall,
\prl 65 (1990) 2939};
for a discussion of the NS terms both in general 
and in the MSSM context see Ref.~\ref\jj{I.~Jack and D.R.T.~Jones,
\plb 457 (1999)  101}, \ref\hethr{J.P.J.~Hetherington,
JHEP  0110 (2001)  024}, and for model-building applications see 
for example Ref.~\ref\bfpt{F. Borzumati et al \npb  555 (1999)  53\semi
P.J.~Fox, A.E.~Nelson and N.~Weiner,
JHEP 0208 (2002) 035
}. For application of NS R-parity violating terms to leptogenesis, 
see Ref.~\ref\HambyeZS{
T.~Hambye, E.~Ma and U.~Sarkar,
\npb 590 (2000)  429
}, and for a review of general soft breaking see 
Ref.~\ref\KongHB{O.C.W.~Kong, hep-ph/0205205}.

In this paper we describe the renormalisation of the 
most general possible softly-broken 
version of the MSSM incorporating both RPV and NS terms. It is interesting 
that, as we shall see, 
with the generalisation to the RPV case the connection between 
the NS terms and cubic scalar interactions involving \sic\ mass terms is 
not universal. 

The unbroken ${\cal N} = 1$ theory is defined 
by the superpotential 
\eqn\superpot{W = W_1 + W_2,}
where 
\eqn\sperrpc{
W_1 =  Y_u Q  u^c H_2 +   Y_d Q d^c  H_1 +  Y_{e} L  e^c  H_1}
 and 
\eqn\sperrpv{
W_2 = \frak{1}{2} (\Lambda_E) e^c L L +  \frak{1}{2}(\Lambda_U) u^c d^c d^c 
+ (\Lambda_D) d^c L Q. }
In these equations, generation $(i,j\cdots)$, $SU_2 (a,b\cdots)$, 
and $SU_3 (\alpha, \beta \cdots)$  indices are 
contracted in ``natural'' fashion from left to right, thus for example
\eqn\indics{
 \Lambda_D d^c L Q 
\equiv  \epsilon_{ab}(\Lambda_D)^{ijk} (d^c)_{i\alpha} L^a_{j} Q^{b\alpha}_k
.}
For the generation indices we indicate complex conjugation by 
lowering the indices, thus $(Y_u)_{ij} = (Y^*_u)^{ij}$. 

We omit possible mass terms $H_1 H_2$ and $L H_2$ because, as we shall 
see, the consequent terms in the Lagrangian will be included as a special 
case of our general structure.

We now add soft-breaking terms as follows:
\eqn\spersspc{\eqalign{
L_1 &= \sum_{\phi}
m_{\phi}^2\phi^*\phi + \left[m_3^2 H_1
H_2 + \sum_{i=1}^3\half M_i\lambda_i\lambda_i  + {\rm h.c. }\right]\cr 
&+ \left[ h_u Q  u^c H_2 +  h_d Q  d^c  H_1 +  h_{e} L  e^c  H_1  
+ {\rm h.c. }\right],\cr
L_2 &= m_R^2   H_1^* L + m_K^2 L H_2 + \frak{1}{2} h_E e^c L L 
+  \frak{1}{2} h_U u^c d^c d^c
+ h_D d^c L Q  + {\rm h.c.,}\cr
L_3 &=  m_4 \psi_{H_1}\psi_{H_2} + 
R_5 H_2^* L e^c +  R_7  H_2^*   Q d^c  +  R_9 H_1^* Q u^c + {\rm h.c.,}\cr
L_4 &= m_r  \psi_{L}\psi_{H_2} +R_1 L^* Q u^c +  R_2 H_1 H_2^* e^c 
+ R_3 u^c e^c d^{c*}
+ \frak{1}{2}R_4 Q Q  d^{c*} + {\rm h.c.}}} 

Thus $L_{1\cdots4}$ correspond to  SRPC, SRPV, NSRPC and NSRPV respectively, 
where $\hbox{SRPC} \equiv \hbox{ Standard R-parity Conserving}$ etc.
All the scalar terms in Eq.~\spersspc\ were first listed 
(as far as we are aware) in Ref.~\HambyeZS.
It is easy to verify that if we set $m_4 = \mu$, $m_r = \kappa$, 
\eqn\susylimit{\eqalign{
(R_1)^{ij}_k &= \kappa_k (Y_u)^{ij}, \quad (R_2)^i = \kappa_j (Y_e)^{ji},\quad 
R_3 = R_4 = 0,\cr
(R_5)^{ij} &= -\mu (Y_e)^{ij} + \kappa_k (\Lambda_E)^{jki},\cr
(R_7)^{ij} &= -\mu (Y_d)^{ij} + \kappa_k (\Lambda_D)^{jki},\cr
(R_9)^{ij} &= \mu (Y_u)^{ij},\cr}}
and
\eqn\susylimitb{\eqalign{
m_1^2 &= \mu^2, \quad m_2^2 = \mu^2 + \kappa^i\kappa_i,\quad 
(m_L^2)^i_j = \kappa^i\kappa_j, \quad (m_R^2)^i  = \mu \kappa^i,\cr
m_3^2 &= h_{u,d,e,U,D,E} =  M_i =  m_Q^2 = m_{e^c}^2 = m_{d^c}^2 = 
m_{u^c}^2 = m_K^2 = 0,}}
then the theory becomes supersymmetric, with 
mass and interaction terms corresponding to the inclusion
in Eqs.~\sperrpc, \sperrpv\ of 
the terms $\mu H_1 H_2$ and
$\kappa L H_2$ respectively. 
(We have assumed for simplicity that $\mu$ is real). This  limiting case 
provides a useful check for our results. 
We have separated the 
soft terms into Eqs.~\susylimit,~\susylimitb\ because those appearing in the 
latter do not contribute to the $\beta$-functions for those 
appearing in the former. Note that, as 
we indicated earlier, we have two interactions ($R_{3,4}$) which 
cannot be generated by \sic\ mass terms.  These interactions violate 
$L, B$ respectively; thus we may expect their 
phenomenological consequences  to be comparable to $h_{E,D}$ 
and $h_U$ respectively.  
In Ref.~\jj\ we gave the general results for the one-loop 
soft $\beta$-functions incorporating NS soft terms, and 
corresponding results for the MSSM in the RPC case. Here 
we generalise the latter results to include RPV, and take the opportunity to 
correct some errors in Ref.~\jj.  

The various one-loop anomalous dimensions were given in, for example, 
Ref.~\ADD; we reproduce them below for convenience (we suppress a $\lf$ loop 
factor throughout):
\eqn\Gresults{\eqalign{
(\ga_L)^i_j &= (Y_e)^{ik}(Y_e)_{jk} + (\Lambda_E)^{kim}(\Lambda_E)_{kjm}
+3(\Lambda_D)^{kim}(\Lambda_D)_{kjm} -2C_H\delta^i_j, \cr
(\ga_{e^c})^i_j &= (Y_e)^{ki}(Y_e)_{kj} + (\Lambda_E)^{ikm}(\Lambda_E)_{jkm}
-2C_{e^c}\delta^i_j, \cr
(\ga_Q)^i_j &= (Y_d)^{im}(Y_d)_{jm} + 
(Y_u)^{im}(Y_u)_{jm}+ (\Lambda_D)^{qmi}(\Lambda_D)_{qmj} -2C_Q\delta^i_j, \cr
(\ga_{d^c})^i_j &= 2(Y_d)^{mi}(Y_d)_{mj} 
+ 2(\Lambda_D)^{ikm}(\Lambda_D)_{jkm} + 2(\Lambda_U)^{kim}(\Lambda_U)_{kjm}
-2C_{d^c}\delta^i_j, \cr
(\ga_{u^c})^i_j &= 2(Y_u)^{mi}(Y_u)_{mj} 
+ (\Lambda_U)^{ikm}(\Lambda_U)_{jkm} 
-2C_{u^c}\delta^i_j, \cr
\ga_{H_1} &= 3(Y_d)^{ij}(Y_d)_{ij} + (Y_e)^{ij}(Y_e)_{ij}-2C_H, \cr
\ga_{H_2} &= 3(Y_u)^{ij}(Y_u)_{ij} -2C_H, \cr
(\ga_{LH_1})^i &=  -3(\Lambda_D)^{kim}(Y_d)_{mk}
-(\Lambda_D)^{kim}(Y_e)_{mk}, \cr}}
where
\eqn\Cdefs{\eqalign{  
C_{Q}&= \frak{4}{3}g_3^2 + \frak{3}{4}g_2^2 +\frak{1}{60}g_1^2,\quad
C_{u^c}= \frak{4}{3}g_3^2 +\frak{4}{15}g_1^2,\quad
C_{d^c}= \frak{4}{3}g_3^2 +\frak{1}{15}g_1^2,\cr
C_{e^c}&= \frak{3}{5}g_1^2,\quad
C_H = \frak{3}{4}g_2^2 +\frak{3}{20}g_1^2.\cr}}

The one loop results for the various $R$-terms follow from Eq.~(2.6) of 
Ref.~\jj\ and are given as follows: 

\eqna\Rresults$$\eqalignno{
\left(\beta_{R_1}\right)^{ij}_k&
=(\ga_L)^m_k(R_1)^{ij}_m+(\ga_Q)^i_m(R_1)^{mj}_k
+(\ga_{u^c})_m^j(R_1)^{im}_k+(\ga_{LH_1})_k(R_9)^{ij}+4C_H(R_1)^{ij}_k\cr
&+6(R_1)^{lm}_k(Y_u)_{lm}(Y_u)^{ij}
+2(Y_e)_{km}(Y_d)^{il}(R_3)^{jm}_l-2(R_3)_n^{jl}(\Lambda_E)_{lkm}
(\Lambda_D)^{nmi}\cr
&-2(\Lambda_D)_{lkm}(\Lambda_D)^{lni}(R_1)^{mj}_n
+4(\Lambda_D)_{lkm}(\Lambda_U)^{jln}(R_4)^{im}_n-2(\Lambda_D)_{mkl}
(Y_u)^{lj}(R_7)^{im}\cr 
&-8(m_r)_kC_H(Y_u)^{ij}+2(R_9)^{lj}(\Lambda_D)_{mkl}(Y_d)^{im}
\cr& 
-4m_4(Y_d)^{im}(Y_u)^{lj}(\Lambda_D)_{mkl}
+4(m_r)_n(Y_u)^{lj}(\Lambda_D)^{mni}(\Lambda_D)_{mkl}, & \Rresults a\cr
\left(\beta_{R_2}\right)^i&
=(\ga_{H_1}+\ga_{H_2})(R_2)^i+(\ga_{e^c})^i_k(R_2)^k
+(\ga_{LH_1})_k(R_5)^{ki}+4C_H(R_2)^i\cr
&+2(R_2)^k(Y_e)^{ji}(Y_e)_{jk}
+6(Y_e)^{ji}(\Lambda_D)_{ljk}(R_7)^{kl}+2(R_5)^{kl}(Y_e)^{ji}
(\Lambda_E)_{ljk}\cr
&+6(Y_d)^{lk}(Y_u)_{lj}(R_3)^{ji}_k
-8(m_r)_j C_H(Y_e)^{ji}, & \Rresults b\cr
\left(\beta_{R_3}\right)^{ij}_k&
=(\ga_{u^c})^i_l(R_3)^{lj}_k+(\ga_{e^c})^j_l(R_3)^{il}_k
+(\ga_{d^c})^l_k(R_3)^{ij}_l +4C_{d^c} (R_3)^{ij}_k \cr 
&+4(R_1)^{mi}_l\left[(Y_d)_{mk} (Y_e)^{lj}+  
(\Lambda_D)_{knm} (\Lambda_E)^{jnl}\right]
+4(R_2)^{j}(Y_d)_{lk}(Y_u)^{li}\cr
&-4 (R_3)^{lj}_m (\Lambda_U)_{lkn} (\Lambda_U)^{imn}
-4(R_5)^{mj}  (Y_u)^{li}(\Lambda_D)_{kml}\cr
&+4(R_9)^{li}  (Y_e)^{mj}(\Lambda_D)_{kml}
-8m_4(Y_u)^{li}(Y_e)^{mj}(\Lambda_D)_{kml}\cr
&-8(m_r)_n(Y_u)^{li}\left[(\Lambda_E)^{jmn}(\Lambda_D)_{kml}
+(Y_d)_{lk}(Y_e)^{nj} \right], 
& \Rresults c\cr
\left(\beta_{R_4}\right)^{ij}_k&
=(\ga_Q)^i_l(R_4)^{lj}_k
+ \frak{1}{2}(\ga_{d^c})^l_k(R_4)^{ij}_l +2C_{d^c} (R_4)^{ij}_k \cr 
&-2(R_1)^{il}_n(\Lambda_D)^{mnj} (\Lambda_U)_{lkm}
-2(R_4)^{il}_m  \left[(Y_d)_{lk}(Y_d)^{jm}
+ (\Lambda_D)_{knl} (\Lambda_D)^{mnj}\right]\cr
&-2(R_7)^{il}  (Y_u)^{jm}(\Lambda_U)_{mkl}
+2(R_9)^{il}  (Y_d)^{jm}(\Lambda_U)_{lkm}
-4m_4(Y_d)^{il}(Y_u)^{jm}(\Lambda_U)_{mkl}\cr
&+4(m_r)_m(Y_u)^{jn}(\Lambda_D)^{lmi}(\Lambda_U)_{nkl} 
+ (i\leftrightarrow j),
& \Rresults d\cr
\left(\beta_{R_5}\right)^{ij}&=
\ga_{H_2}(R_5)^{ij} + (\ga_L)^i_k (R_5)^{kj}
+ (R_5)^{ik} (\ga_{e^c})^j_k
+4 C_H \left[(R_5)^{ij} + 2 m_4 (Y_e)^{ij}\right]
& \cr 
&+(\ga_{LH_1})^iR_2^j +6R_7^{kl}(Y_d)_{kl}(Y_e)^{ij}+2R_5^{kl}(Y_e)_{kl}(Y_e)^{ij}\cr
&-6R_7^{kl}(\Lambda_D)_{lmk}(\Lambda_E)^{jim}+2R_5^{kl}(\Lambda_E)_{lkm}
(\Lambda_E)^{jim}& \cr 
&-2R_2^l(Y_e)_{kl}(\Lambda_E)^{jik}
-6(\Lambda_D)^{lik}(Y_u)_{km}(R_3)_l^{mj}
+8(m_r)_kC_H(\Lambda_E)^{jik}, &\Rresults e\cr
\left(\beta_{R_7}\right)^{ij} 
&= \ga_{H_2}(R_7)^{ij} + (\ga_Q)^i_k (R_7)^{kj} 
+ (R_7)^{ik} (\ga_{d^c})^j_k
+4 C_H \left[(R_7)^{ij} + 2 m_4 (Y_d)^{ij}\right] 
& \cr
&+ 6(R_7)^{mn}(Y_d)_{mn}(Y_d)^{ij}
+2(R_5)^{mn}(Y_e)_{mn}(Y_d)^{ij} 
+6R_7^{kl}(\Lambda_D)_{lmk}(\Lambda_D)^{jmi} & \cr
&-2R_5^{lm}(\Lambda_E)_{mlk}(\Lambda_D)^{jki}
- 2(R_7)^{kj}(Y_u)_{kl}(Y_u)^{il}
+2(R_9)^{ik}(Y_u)_{lk}(Y_d)^{lj}& \cr
&-2(R_1)^{im}_l(Y_u)_{km}(\Lambda_D)^{jlk} 
+ 2(R_2)^{k}(Y_e)_{lk}(\Lambda_D)^{jli}
+ 4(R_4)^{ik}_l (\Lambda_U)^{mjl}(Y_u)_{km} & \cr
&- 4m_4(Y_u)^{il}(Y_u)_{kl}(Y_d)^{kj}
+ 4(m_r)_l (Y_u)^{im}(\Lambda_D)^{jlk}(Y_u)_{km}\cr
&-8 (m_r)_k C_H (\Lambda_D)^{jki}, & \Rresults f\cr
\left(\beta_{R_9}\right)^{ij}&= 
\ga_{H_1}(R_9)^{ij} + (\ga_Q)^i_k (R_9)^{kj} 
+ (R_9)^{ik} (\ga_{u^c})^j_k
+4 C_H \left[(R_9)^{ij} - 2 m_4 (Y_u)^{ij}\right] 
& \cr
& +(\ga_{L H_1})^k (R_1)^{ij}_k 
+ 6(R_9)^{mn}(Y_u)_{mn}(Y_u)^{ij} - 2(R_9)^{mj}(Y_d)_{mn}(Y_d)^{in}& \cr
&+2(R_7)^{im}(Y_d)_{nm}(Y_u)^{nj} 
+ 4(R_4)^{ip}_m (\Lambda_U)^{jmn}(Y_d)_{pn} +
2(R_3)^{jp}_m (Y_e)_{kp}\Lambda^{mki}_D& \cr 
&+2(R_1)^{mj}_p (\Lambda_D)^{npi} (Y_d)_{mn}
  + 4m_4(Y_d)^{ik}(Y_d)_{lk}(Y_u)^{lj}& \cr
&- 4(m_r)_l (\Lambda_D)^{mli}(Y_d)_{nm}(Y_u)^{nj}. & \Rresults g\cr
}$$

The one loop results for the various $\phi\phi^*$ mass-terms 
follow from Eq.~(2.7c) of 
Ref.~\jj: 
\eqna\Mresults$$\eqalignno{
(\beta_{m_Q^2})^i_j &=  
(m_Q^2)^i_k\left[ (Y_u)^{kl}(Y_u)_{jl}+(Y_d)^{kl}(Y_d)_{jl} 
+(\Lambda_D)^{mlk}(\Lambda_D)_{mlj}\right]\cr
&+(m_Q^2)_j^k\left[ (Y_u)_{kl}(Y_u)^{il}+(Y_d)_{kl}(Y_d)^{il} 
+(\Lambda_D)_{mlk}(\Lambda_D)^{mli}\right]\cr
&+2\bigl[m_1^2(Y_d)^{il}(Y_d)_{jl}+
m_2^2(Y_u)^{il}(Y_u)_{jl}
+(m_{u^c}^2)^k_l(Y_u)^{il}(Y_u)_{jk}\cr&
+(m_{d^c}^2)^k_l\left[(Y_d)^{il}(Y_d)_{jk}+
(\Lambda_D)^{lmi}(\Lambda_D)_{kmj}\right]+
(\Lambda_D)^{mli}(\Lambda_D)_{mkj}(m_L^2)^k_l\bigr]\cr
&-2(m_R^2)_k(Y_d)_{jl}(\Lambda_D)^{lki} 
-2(m_R^2)^k(Y_d)^{il}(\Lambda_D)_{lkj}\cr 
&+2\left[(h_u)^{ik}(h_u)_{jk}+(h_d)^{ik}(h_d)_{jk}
+(h_D)^{kli}(h_D)_{klj}\right]\cr
&
+2\left[(R_1)^{ik}_l (R_1)^l_{jk}+(R_7)^{ik} (R_7)_{jk}+
(R_9)^{ik} (R_9)_{jk}+2(R_4)^{ik}_l (R_4)^l_{jk}\right]\cr
&
-4\bigl[m_4^2\left[ (Y_u)^{ik}(Y_u)_{jk}+(Y_d)^{ik}(Y_d)_{jk}\right]
+(m_r)^l (m_r)_l (Y_u)^{ik}(Y_u)_{jk}\cr&
+(m_r)^m (m_r)_l (\Lambda_D)^{kli}(\Lambda_D)_{kmj}\bigr]
+4m_4\left[(\Lambda_D)^{kli}(Y_d)_{jk}(m_r)_l+
(\Lambda_D)_{klj}(Y_d)^{ik}(m_r)^l \right]\cr&
-\left[\frak{32}{3}g_3^2M_3^2+6g_2^2M_2^2
+\frak{2}{15}g_1^2M_1^2 -\frak{1}{5}g_1^2\Scal\right]\delta^i_j,
& \Mresults a\cr
(\beta_{m_{u^c}^2})^i_j &= (m_{u^c}^2)^i_m\left[2(Y_u)^{lm}(Y_u)_{lj}
+ (\Lambda_U)^{mkl}(\Lambda_U)_{jkl}\right] 
+ 4(m_{Q}^2)^k_l(Y_u)^{li}(Y_u)_{kj}\cr
&+(m_{u^c}^2)^m_j\left[2(Y_u)_{lm}(Y_u)^{li}
+ (\Lambda_U)_{mkl}(\Lambda_U)^{ikl}\right]
+ 4(m_{2}^2)(Y_u)^{li}(Y_u)_{lj}\cr
&+ 4(m_{d^c}^2)^k_m  (\Lambda_U)^{iml}(\Lambda_U)_{jkl}
+4(h_u)^{ki}(h_u)_{kj}
+2(h_U)^{ikl}(h_U)_{jkl}\cr&
+4\left[(R_1)^{ki}_l (R_1)^l_{kj}+(R_9)^{ki} (R_9)_{kj} 
\right]+2(R_3)^{ik}_l(R_3)^l_{jk} 
-8m_4^2 (Y_u)^{li}(Y_u)_{lj} \cr&
-8(m_r)^k (m_r)_k (Y_u)^{li}(Y_u)_{lj}
-\left[\frak{32}{3}g_3^2M_3^2+\frak{32}{15}g_1^2M_1^2
+\frak{4}{5}g_1^2\Scal\right]\delta^i_j,
& \Mresults b\cr
(\beta_{m_{d^c}^2})^i_j &= 2(m_{d^c}^2)^i_m\left[(Y_d)^{lm}(Y_d)_{lj}
+ (\Lambda_U)^{kml}(\Lambda_U)_{kjl}  
+  (\Lambda_D)^{mkl}(\Lambda_D)_{jkl}\right] 
\cr
&+2(m_{d^c}^2)^m_j\left[(Y_d)_{lm}(Y_d)^{li}
+ (\Lambda_U)_{kml}(\Lambda_U)^{kil}
+  (\Lambda_D)_{mkl}(\Lambda_D)^{ikl}\right]
\cr
& + 4(m_{Q}^2)^k_l(Y_d)^{li}(Y_d)_{kj} 
+4  (\Lambda_U)^{kil}\left[(m_{u^c}^2)^m_k(\Lambda_U)_{mjl}
+ (m_{d^c}^2)^m_l(\Lambda_U)_{kjm}\right]
\cr&
+ 4m_{1}^2(Y_d)^{li}(Y_d)_{lj}
+ 4(\Lambda_D)^{ikl}\left[(m_{L}^2)^m_k (\Lambda_D)_{jml}
+(m_{Q}^2)^m_l (\Lambda_D)_{jkm}\right]\cr
&
-4(m_R^2)_k(Y_d)_{lj}(\Lambda_D)^{ikl} 
-4(m_R^2)^k(Y_d)^{li}(\Lambda_D)_{jkl}\cr 
&+4\left[(h_d)^{ki}(h_d)_{kj}
+(h_U)^{kil}(h_U)_{kjl} + (h_D)^{ikl}(h_D)_{jkl}\right]\cr&
+2\left[(R_3)^{i}_{kl} (R_3)^{kl}_{j}+
2(R_4)^{i}_{kl} (R_4)^{kl}_{j}
+2(R_7)^{ki} (R_7)_{kj}\right] -8m_4^2 (Y_d)^{ki}(Y_d)_{kj} \cr&
-8(m_r)^k (m_r)_l  (\Lambda_D)^{ilm}(\Lambda_D)_{jkm}
+8(m_r)^l m_4  (Y_d)^{ki}(\Lambda_D)_{jlk}\cr&
+8(m_r)_l m_4  (Y_d)_{kj}(\Lambda_D)^{ilk}
-\left[\frak{32}{3}g_3^2M_3^2+\frak{8}{15}g_1^2M_1^2
-\frak{2}{5}g_1^2\Scal\right]\delta^i_j,
& \Mresults c\cr
(\beta_{m_{L}^2})^i_j &= 
(m_{L}^2)^i_m\left[(Y_e)^{mk}(Y_e)_{jk}
+(\Lambda_E)^{kml}(\Lambda_E)_{kjl}
+ 3(\Lambda_D)^{kml}(\Lambda_D)_{kjl}\right]\cr
&+(m_{L}^2)^m_j\left[(Y_e)_{mk}(Y_e)^{ik}
+(\Lambda_E)_{kml}(\Lambda_E)^{kil}
+ 3(\Lambda_D)_{kml}(\Lambda_D)^{kil}\right]\cr 
&+2m_1^2 (Y_e)^{ik}(Y_e)_{jk}
-(m_R^2)^i(Y_e)^{lk}(\Lambda_E)_{kjl}
-(m_R^2)_j(Y_e)_{lk}(\Lambda_E)^{kil}\cr&
+2(m_R^2)^k(Y_e)^{il}(\Lambda_E)_{ljk}
+2(m_R^2)_k(Y_e)_{jl}(\Lambda_E)^{lik}
-3(m_R^2)^i(Y_d)^{kl}(\Lambda_D)_{ljk}\cr&
-3(m_R^2)_j(Y_d)_{kl}(\Lambda_D)^{lik}
+2(m_{e^c}^2)^k_l\left[(Y_e)^{il}(Y_e)_{jk}
+(\Lambda_E)^{lim}(\Lambda_E)_{kjm}\right] \cr
&+2(m_{L}^2)^m_l(\Lambda_E)^{kil}(\Lambda_E)_{kjm}
+6(\Lambda_D)^{kim}\left[(m_{Q}^2)^l_m (\Lambda_D)_{kjl}
+(m_{d^c}^2)^l_k (\Lambda_D)_{ljm}\right]\cr&
+2(h_e)^{ik}(h_e)_{jk}
+2(h_E)^{kil}(h_E)_{kjl} +6(h_D)^{kil}(h_D)_{kjl}
+6(R_1)^{i}_{kl} (R_1)^{kl}_{j}\cr& + 2(R_5)^{ik} (R_5)_{jk}
-4\bigl[ m_4^2(Y_e)^{ik}(Y_e)_{jk}
+(m_r)^k (m_r)_l (\Lambda_E)^{mil}(\Lambda_E)_{mjk}\cr&
 +(m_r)^k  m_4 (Y_e)^{il}(\Lambda_E)_{ljk}
+(m_r)_k  m_4 (Y_e)_{jl}(\Lambda_E)^{lik}\bigr]\cr
&-8(m_r)^i (m_r)_j C_H -\left[6g_2^2M_2^2
+\frak{6}{5}g_1^2M_1^2+\frak{3}{5}g_1^2\Scal\right]\delta^i_j,
& \Mresults d\cr
(\beta_{m_{e^c}^2})^i_j &= 
(m_{e^c}^2)^i_m\left[2(Y_e)^{km}(Y_e)_{kj}
+(\Lambda_E)^{mkl}(\Lambda_E)_{jkl}
\right]\cr
&+(m_{e^c}^2)^m_j\left[2(Y_e)_{km}(Y_e)^{ki}
+(\Lambda_E)_{mkl}(\Lambda_E)^{ikl}\right] 
+4m_1^2 (Y_e)^{ki}(Y_e)_{kj}\cr&
+4(m_R^2)^k(Y_e)^{mi}(\Lambda_E)_{jmk}
+4(m_R^2)_k(Y_e)_{mj}(\Lambda_E)^{imk}
\cr&
+4(m_{L}^2)^k_l\left[(Y_e)^{li}(Y_e)_{kj}
+(\Lambda_E)^{ilm}(\Lambda_E)_{jkm}\right] 
+4(h_e)^{ki}(h_e)_{kj}\cr& 
+2(h_E)^{ikl}(h_E)_{jkl}
+4(R_2)^{i} (R_2)_{j} + 4(R_5)^{ki} (R_5)_{kj}
+6(R_3)^{ki}_l (R_3)_{kj}^l\cr& 
-8\bigl[ m_4^2(Y_e)^{ki}(Y_e)_{kj}
+(m_r)^k (m_r)_l (\Lambda_E)^{iml}(\Lambda_E)_{jmk}\
 +(m_r)^k  m_4 (Y_e)^{li}(\Lambda_E)_{jlk}\cr&
+(m_r)_k  m_4 (Y_e)_{lj}(\Lambda_E)^{ilk}
+  (Y_e)^{ki} (Y_e)_{lj}(m_r)^l (m_r)_k \bigr] 
-\left[\frak{24}{5}g_1^2M_1^2 -\frak{6}{5}g_1^2\Scal\right]\delta^i_j, 
&  \Mresults e\cr
\beta_{m_{1}^2} &= 
2m_{1}^2\left[(Y_e)^{ij}(Y_e)_{ij}
+3(Y_d)^{ij}(Y_d)_{ij}\right]
+2(m_{L}^2)^k_i(Y_e)^{ij}(Y_e)_{kj}
+2(m_{e^c}^2)^j_k(Y_e)^{ik}(Y_e)_{ij}\cr&
+6(m_{Q}^2)^j_i(Y_d)^{ik}(Y_d)_{jk}
+6(m_{d^c}^2)^j_k(Y_d)^{ik}(Y_d)_{ij}
+2(h_e)^{ij}(h_e)_{ij}
+6(h_d)^{ij}(h_d)_{ij}\cr&+2(R_2)^{i} (R_2)_{i}+ 6(R_9)^{ij} (R_9)_{ij}
 -(m_R^2)^j (\Lambda_E)_{ljk} (Y_e)^{kl}
-(m_R^2)_j (\Lambda_E)^{ljk} (Y_e)_{kl}\cr&
 -3(m_R^2)^j (\Lambda_D)_{ljk} (Y_d)^{kl}
-3(m_R^2)_j (\Lambda_D)^{ljk} (Y_d)_{kl}-8m_4^2 C_H\cr&
-4(m_r)^i (m_r)_j (Y_e)^{jk}(Y_e)_{ik}-
\left[6g_2^2M_2^2+\frak{6}{5}g_1^2M_1^2+\frak{3}{5}g_1^2\Scal\right],
& \Mresults f\cr
\beta_{m_{2}^2} &=
6\left[m_{2}^2(Y_u)^{ij}(Y_u)_{ij}
+(m_{Q}^2)^i_j (Y_u)^{jk}(Y_u)_{ik} +(m_{u^c}^2)^i_j (Y_u)^{kj}(Y_u)_{ki}  
\right]\cr&
+6(h_u)^{ij}(h_u)_{ij}+2(R_2)^{i} (R_2)_{i}+ 2(R_5)^{ij} (R_5)_{ij}
+ 6(R_7)^{ij} (R_7)_{ij}\cr&
-8C_H \left[m_4^2 + (m_r)^i(m_r)_i\right]
-\left[6g_2^2M_2^2+\frak{6}{5}g_1^2M_1^2-\frak{3}{5}g_1^2\Scal\right],
& \Mresults g\cr
(\beta_{m_R^2})^i &= -m_{1}^2 (Y_e)_{jl} (\Lambda_E)^{lij}
-(m_L^2)^i_m(\Lambda_E)^{lmj}(Y_e)_{jl} &\cr 
&-2(\Lambda_E)^{kij}\left[(Y_e)_{nk}(m_L^2)^n_j 
+ (Y_e)_{lj}(m_{e^c}^2)^l_k\right] \cr&+(m_R^2)^k\left[ 
(\Lambda_E)^{mij}(\Lambda_E)_{mkj} - (Y_e)_{km}(Y_e)^{im}
+3(\Lambda_D)^{mij}(\Lambda_D)_{mkj}\right]\cr
&-3(Y_d)_{jk}\left[m_{1}^2(\Lambda_D)^{kij} + (m_L^2)^i_l(\Lambda_D)^{klj}
+2 (m_Q^2)^j_l(\Lambda_D)^{kil} + 2(m_{d^c}^2)^k_l(\Lambda_D)^{lij}\right]\cr
& +(m_R^2)^i\left[
(Y_e)^{km}(Y_e)_{km} + 3(Y_d)^{km}(Y_d)_{km}\right]
-2h_E^{kij}(h_e)_{jk}-6h_D^{kij}(h_d)_{jk}\cr
&+ 6(R_1)^i_{jk}(R_9)^{jk} +2 (R_5)^{ij} (R_2)_j -8m_4(m_r)^i C_H\cr
&
+4m_4(m_r)^l (Y_e)^{in}(Y_e)_{ln}+4(m_r)_j(m_r)^l (\Lambda_E)^{nij}(Y_e)_{ln},
& \Mresults h\cr
}$$
where
\eqn\scaldef{
\Scal = m_2^2 - m_1^2 + \Tr\left[m_Q^2 + m_{d^c}^2+ m_{e^c}^2
-  m_L^2 - 2 m_{u^c}^2\right].}
$\Scal$ arises as a renormalisation of the 
Fayet-Iliopoulos $D$-term. It is one-loop RG invariant in the absence of
NS  terms, and is then small at all relevant scales if it is  zero
at gauge unification; however  this RG invariance no longer holds in the
presence of NS terms\ref\jjfi{I.~Jack and D.R.T.~Jones,
\prd 63 (2001) 075010 
}.

For the $\phi\phi$-type terms:
\eqna\MMresults$$\eqalignno{
\beta_{m_3^2} &= \left(\ga_{H_1} + \ga_{H_2}\right)m_3^2  
+(\ga_{L H_1})_i (m_K^2)^i - 2(R_5)_{ij} (h_e)^{ij}\cr& 
-6 (R_7)_{ij} (h_d)^{ij}
+6 (R_9)_{ij} (h_u)^{ij} + m_4(6g_2^2M_2+\frak{6}{5}g_1^2M_1),
& \MMresults a\cr  
\beta_{m_K^2}^i &=(\ga_L)^i_j (m_K^2)^j + \ga_{H_2}  (m_K^2)^i
+ (\ga_{L H_1})^i m_3^2 +  6(R_1)^i_{jk}(h_u)^{jk}\cr& 
+ 2(R_2)_j  (h_e)^{ij} +  2(R_5)_{jk}  (h_E)^{kij}
+ 6 (R_7)_{jk} (h_D)^{kij}\cr&
+(m_r)^i \left( 6g_2^2M_2+\frak{6}{5}g_1^2M_1\right),
& \MMresults b\cr 
}$$
and for the $\psi\psi$ terms:
\eqna\psimfour$$\eqalignno{
\beta_{m_4} &= \left(\ga_{H_1} + \ga_{H_2}\right) m_4 
+ (m_r)^{i}  (\ga_{L H_1})_i & \psimfour a\cr
(\beta_{m_r})^i &=   (m_r)^{i}\ga_{H_2} 
+  (m_r)^{j}(\ga_{L})^i_j + m_4  (\ga_{L H_1})^i. & \psimfour b\cr}$$
In the \sic\ limit described in Eqs.~\susylimit,\susylimitb, 
Eqs.~\psimfour{}\ become the $\beta$-functions for the 
corresponding \sic\ mass terms. As indicated earlier, this limit gives 
a useful check on all our results. A further check is provided 
by the fact that, as often discussed in the literature (for example 
Refs.~\ADD,\deCarlosDU,\KongHB) in the presence of general 
R-parity violation   the distinction between the 
lepton doublets $L_i$ 
and the Higgs doublet $H_1$ is artificial. This means that by, for example, 
``promoting'' $m_L^2$ to be a $4\times4$ matrix, we can extract from 
Eq.~\Mresults{d}\ the results for both $\beta_{m_1^2}$ 
(Eq.~\Mresults{f}) and $\beta_{m_R^2}$ (Eq.~\Mresults{h}). 
In general our results reduce to and agree with those of Ref.~\ADD\ 
when all NS terms are removed, up to very minor typos.

In the unbroken theory,  the simplified case is often considered   when
each dimensionless  coupling matrix is assumed to have only one non-zero
entry: $(Y_u)^{33} = \lambda_t, (Y_d)^{33} = \lambda_b,  
(Y_e)^{33} = \lambda_{\tau}, 
(\Lambda_{E})^{323} = \lambda,  (\Lambda_{D})^{333} =
\lambda^{'},  (\Lambda_{U})^{323} = \lambda^{''}$ (see for example 
Ref.~\ref\AnanthanarayanPC{
B.~Ananthanarayan and P.~N.~Pandita,
\prd 63 (2001)  076008 
}\ for an analysis of the associated infrared fixed point structure, and 
Ref.~\ref\MambriniSJ{
Y.~Mambrini and G.~Moultaka,
\prd 65 (2002)  115011}\ for a more general discussion). 
This is evidently phenomenologically 
sensible for $Y_{u,d,e}$,  but is less obviously 
justifiable for $\Lambda_{E,D,U}$. 
Moreover 
this set of couplings is not closed under renormalisation
\ref\ADDb{B.C.~Allanach, A.~Dedes and H.K.~Dreiner,
\prd 60 (1999)  056002
}; at one loop a minimal set
that {\it is\/}  would also 
include  $(Y_e)^{23}$ and $(\Lambda_{D})^{323}$.

In the R-parity conserving case when $\Lambda_{E,D,U} = 0$ 
the single generation 
approximation does close under renormalisation and is naturally extended to the 
soft breaking case by setting $(R_9)^{33} = r_9 = m_9\lambda_t$,
$(R_7)^{33} = r_7 = -m_7\lambda_b$ and 
$(R_5)^{33} = r_5 = -m_5\lambda_{\tau}$, 
the signs being chosen so that the supersymmetric fixed point
corresponds to  $m_4 = m_5 = m_7 = m_9$ (see Eq.~\susylimit). 
We then obtain: 
\eqna\newres$$\eqalignno{
\beta_{m_4} &= (\lambda_{\tau}^2 + 3 \lambda_b^2 
+ 3\lambda_t^2- 4 C_H)m_4
, &\newres a\cr
\beta_{m_5} &= 
(\lambda_{\tau}^2 - 3 \lambda_b^2 + 3\lambda_t^2)m_5
+6m_7\lambda_b^2 +
(4m_5 - 8m_4)C_H, &\newres b\cr
\beta_{m_7} &= (-\lambda_{\tau}^2 + 3 \lambda_b^2 + \lambda_t^2)m_7
+2m_5\lambda_{\tau}^2 +2\lambda_t^2 (2m_4-m_9)\cr & + 
(4m_7 - 8m_4)C_H, &\newres c\cr
\beta_{m_9} &= (\lambda_{\tau}^2 + \lambda_b^2 + 3\lambda_t^2)m_9
-2m_7\lambda_b^2 +4m_4\lambda_b^2 +
(4m_9 - 8m_4)C_H. &\newres d\cr}$$
Eqs.~\newres{a,b}\ above agree with Eq.~(3.5) of Ref.~\jj, but 
Eqs.~\newres{c,d}\ differ. The error made in Ref.~\jj\ was neglecting 
possible minus signs associated with $SU_2$ contractions 
involving $\epsilon^{ab}$ (in the RPV case 
one must be similarly careful with 
regard to the $SU_3$ tensor $\epsilon^{\alpha\beta\gamma}$).
The analysis of the fixed point 
$m_4 = m_5 = m_7 = m_9$ given in section~4 of Ref.~\jj\ is 
changed somewhat. The stability matrix for the 
evolution of $\frakk{m_5}{m_4}$, $\frakk{m_7}{m_4}$
and $\frakk{m_9}{m_4}$ is given by:

\eqn\matx{
S = \pmatrix{8C_H-6\lambda_b^2 & 6\lambda_b^2 & 0\cr
2\lambda_{\tau}^2 & 8C_H - 2\lambda_{\tau}^2
- 2\lambda_t^2 & - 2\lambda_t^2 \cr
0 & -2\lambda_b^2 & 8C_H - 2\lambda_b^2\cr}}
which has eigenvalues
$8C_H, 8C_H + \Lambda_{1,2}$
where $\Lambda_{1,2}$ are the roots of the quadratic
\eqn\qudrtc{
\Lambda^2 + 2(\lambda_t^2 + \lambda_{\tau}^2 + 4\lambda_b^2)\Lambda
+4(3\lambda_b^2 + 3 \lambda_t^2 + \lambda_{\tau}^2)\lambda_b^2=0.}

Near the quasi-infra-red fixed point (QIRFP) for $\lambda_t$, 
$\lambda_t(M_Z) \approx 1.1$ (corresponding to $\tan\beta \approx 1.7$), 
we can neglect $\lambda_b$ and $\lambda_{\tau}$,
and it is easy to see that our fixed point is stable. 
In the trinification region such that 
$\lambda_t (M_U) \approx \lambda_b(M_U) \approx  
\lambda_{\tau}(M_U)\approx 0.6$, the eigenvalues of $S$ are 
all positive at unification but $8C_H + \Lambda_{1,2}$ are both negative 
at $M_Z$. Thus in this case we would expect substantial deviation from
the supersymmetric limit for $m_{4,\cdots 9}$.

Returning to the NSRPV terms, as an example of their possible effect  we
will investigate the effect of a
nonzero $(R_4)^{33}_3 = r_4$ only; thus for this exercise we will 
assume $\Lambda_{U,D,E} = 0$ as well as all the other $R$-couplings, 
so that baryon number is violated but not lepton number. 
The $R_4$ interactions are similar to
$\Lambda_U$ in violating  baryon number by one unit; but of course $R_4$
involves only sparticles and  so we could expect the upper limit on any
particular component  to be less severe than the limit on a
corresponding component of $\Lambda_U$. For interactions other than  
$(\Lambda_U)_{121}$ and $(\Lambda_U)_{131}$ these 
limits are not very
strict  in any case
\ref\GoityDQ{J.~L.~Goity and M.~Sher,
\plb 346 (1995)  69
[Erratum-ibid. {\bf B}385 (1996)  500]\semi
B.C.~Allanach, A.~Dedes and H.K.~Dreiner,
\prd 60  (1999) 075014
}; 
so if we assume no flavour mixing then $r_4 (M_Z)$
could be as large as  the susy scale.
Therefore even if direct detection of the interaction  is difficult
it will influence the sparticle spectrum via Renormalisation Group evolution. 

Since $r_4$ only contributes to  the $\beta$-functions for $(m_Q^2)^3_3$
and $(m_{d^c}^2)^3_3$   we may expect  the main effect to be on the 3rd
generation squark  masses. 
As an example of its effect, in Figure~1 we plot the light stop mass 
against $r_4(M_Z)$ for the SPS5 benchmark point\ref\AllanachNJ{
B.C.~Allanach {\it et al.},
Eur.\ Phys.\ J.\ C {\bf 25}, 113 (2002)
}.  The SPS5 point is characterised by the fact that one of the 
stop masses is rather light and sensitive to small changes in 
input parameters such 
as the top quark mass. We use the one-loop $\beta$ function for $r_4$, 
two loop $\beta$-functions 
for all other couplings and masses (including $r_4$ only at one loop), 
and  adjust input parameters according to
the supersymmetric spectrum in order to account for
threshold corrections in the manner of
Ref.~\ref\pbmz{
D.M.~Pierce, J.A.~Bagger, K.T.~Matchev and R.J.~Zhang,
\npb491 (1997) 3
}. Our two-loop result for the light stop mass at $r_4(M_Z)=0$ is 
now $257\GeV$, a change from that reported in 
Ref.~\ref\JackSX{
I.~Jack, D.R.T.~Jones and A.F.~Kord,
\plb 579 (2004)  180}, due to use of the exact rather than the 
approximate form of the stop mass matrix from Ref.~\pbmz.
We see that as $r_4(M_Z)$ approaches $0.5\TeV$ 
(corresponding to a value at gauge unification $r_4 \approx 0.3\TeV$) 
the light stop mass varies quite significantly.    
\vskip 0.7cm
\epsfysize= 4in
\centerline{\epsfbox{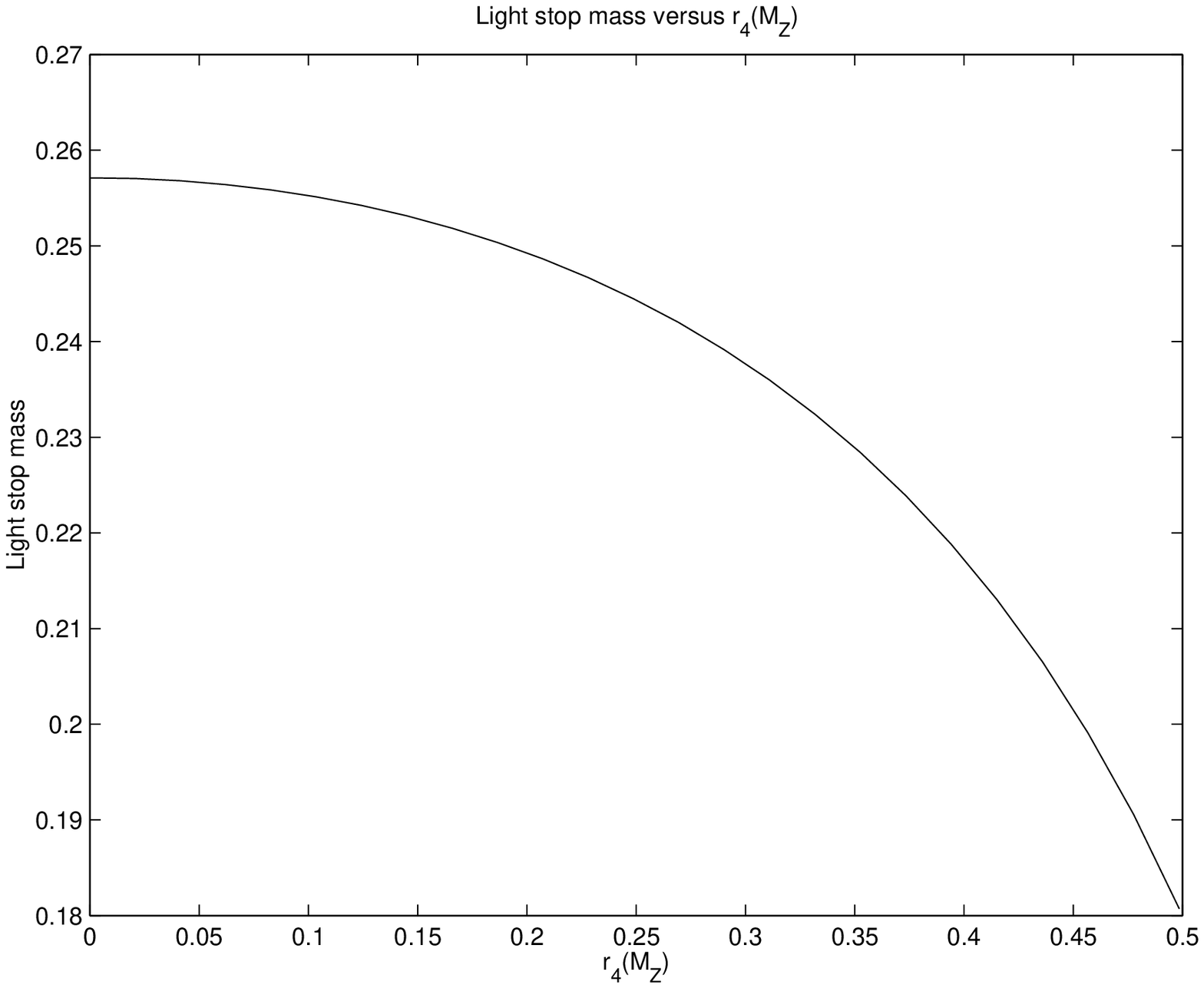}}
\inparg
{\it\noindent{Figure~1: The light stop mass (in TeV) as a 
function of $r_4 (M_Z)$ 
(also in TeV) for the SPS5 Benchmark Point}}
\medskip
\outparg

In conclusion: we have presented the one-loop renormalisation of the 
R-parity violating extension of the MSSM 
with the {\it most general possible\/} set 
of soft breaking terms consistent with naturalness. 
If the amount of flavour mixing is small then the effect of 
non-standard soft R-parity violating terms on the sparticle spectrum 
might be considerable.

\bigskip\centerline{{\bf Acknowledgements}}

DRTJ was  supported by a PPARC Senior Fellowship and a 
CERN Research Associate-ship.  AK was
supported by an Iranian Government Studentship. We also thank Neil Pomeroy,
who participated in the early stages of this investigation, and Otto Kong 
and Ben Allanach for correspondence.

\listrefs
\end